\documentclass[12pt,preprint]{aastex}

\def\h2{H$_{2}$}

\def\n01{$N_{01}$}
\def\t01{$T_{01}$}

\input epsf

\begin{document}

\title{THE MULTI-PHASE INTERGALACTIC MEDIUM TOWARDS PKS 2155-304 }

\author{J. MICHAEL SHULL\altaffilmark{1,2},
JASON TUMLINSON\altaffilmark{3}, and MARK L. GIROUX\altaffilmark{4}
} 
\altaffiltext{1}{CASA, Department of Astrophysical and Planetary Sciences, 
   University of Colorado, Boulder, CO 80309 (mshull@casa.colorado.edu)}
\altaffiltext{2}{Also at JILA, University of Colorado and National Institute 
   of Standards and Technology}
\altaffiltext{3}{Department of Astronomy and Astrophysics, University 
   of Chicago, Chicago, IL 60637 (tumlinso@oddjob.uchicago.edu)}
\altaffiltext{4}{Department of Physics and Astronomy, Box 70652,
   East Tennessee State University, Johnson City, TN 37614 (giroux@polar.etsu.edu)}

\begin{abstract}

We study the cluster of H~I and O~VI absorption systems and the 
claimed detection of O~VIII absorption from the intergalactic medium 
at $z \approx 0.0567$, associated with a group of galaxies toward 
the BL~Lac object PKS~2155-304. As measured by spectrographs on the 
{\it Hubble Space Telescope}, {\it Far Ultraviolet Spectroscopic Explorer}, 
and {\it Chandra}, this system appears to contain gas at a variety of 
temperatures.  We analyze this multi-phase gas in a clumpy-infall model. 
From the absence of C~IV and Si~III absorption in the Ly$\alpha$ clouds, 
we infer metallicities less than 2.5--10\% of solar values.  The only 
metals are detected in two O~VI absorption components, offset by 
$\pm 400$ km~s$^{-1}$ from the group barycenter ($cz \approx 16,600$ 
km~s$^{-1}$).  The \ion{O}{6} components may signify ``nearside" and 
``backside" infall into the group potential well, which coincides with 
the claimed O~VIII absorption. 
If the claimed O~VIII detection is real, our analysis suggests that
clusters of strong Ly$\alpha$ and O~VI absorbers, 
associated with groups of galaxies, may be the ``signposts" of 
shock-heated, metal-enriched baryons.  Through combined UV and X-ray
spectra of H~I and O~VI, VII, and VIII, one may be able to clarify  
the heating mechanism of this multiphase gas.  
\end{abstract}


\section{INTRODUCTION}

With the development of high-throughput ultraviolet and X-ray spectrographs,
we can begin to account for a significant fraction of the low-redshift 
``missing baryons" (Fukugita, Hogan, \& Peebles 1998; Shull 2003) in the 
intergalactic medium (IGM).  
Recent surveys by the {\it Hubble Space Telescope} (HST) and the
{\it Far Ultraviolet Spectroscopic Explorer} (FUSE) show that $32\pm6$\% 
of the baryons reside in the warm ($10^4$~K) Ly$\alpha$ forest (Penton, Stocke,
\& Shull 2003), while $5-10$\% lie in the shock-heated ($10^{5-6}$~K) 
IGM traced by O~VI absorption (Tripp, Savage, \& Jenkins 2000).    
A predicted 30-40\% of the IGM remains to be found in even hotter gas 
($10^{6-7}$~K) through X-ray absorption lines of O~VII, O~VIII, and perhaps
Ne~IX (Fang et al.\ 2002; Nicastro et al.\ 2002).  

The total baryon density is known to 10\% accuracy from measurements
of light-element nucleosynthesis (D/H) and acoustic oscillations 
in the cosmic microwave background (CMB).  Recent estimates from these
techniques yield consistent values for $\Omega_b$, the fractional 
contribution of baryons to the closure density:   
$\Omega_b = (0.020 \pm 0.002) h^{-2}$ (D/H -- Burles \& Tytler 1998) 
and $\Omega_b = (0.0224 \pm 0.0009) h^{-2}$ (CMB -- Spergel et al.\ 2003; 
Netterfield et al.\ 2002) for a Hubble constant 
$H_0 = 100h$ km~s$^{-1}$~Mpc$^{-1}$. Even $H_0$ has become a
well-measured parameter, with values of $h = 0.71 \pm 0.08$
(Freedman et al.\ 2001) from the Cepheid Key Project with HST. 
Using the CMB values, the co-moving baryon density 
is $\rho_b = (4.2 \pm 0.2) \times 10^{-31}$ g~cm$^{-3}$ 
and the hydrogen density is $n_H = (1.90 \pm 0.2) \times 10^{-7}$
cm$^{-3}$ for primordial helium abundance $Y_p = 0.244$ by mass.   

Numerical simulations of the low-$z$ IGM  (Cen \& Ostriker 1999; 
Dav\'e et al.\ 2001) predict that the gas is distributed nearly 
equally ($30 \pm 10$\% each) in three ``phases": 
(1) warm photoionized gas ($10^4$~K); (2) warm/hot shocked gas 
($10^{5-7}$~K); and (3) collapsed halos, galaxies, and clusters with much 
hotter gas ($T > 10^7$~K).  The warm phase is observable through UV absorption 
lines of H~I, C~III, C~IV, Si~III, and Si~IV, while the warm/hot phase
is detectable in O~VI (UV) and in X-ray lines of O~VII, O~VIII, 
and Ne~IX.   The warm-hot intergalactic medium (WHIM) is thought to be 
produced  by shock-heating during gravitational infall into dark-matter 
filaments co-evolving with the rise in cosmic metallicity.
These heavy elements are theorized to be expelled from galaxies 
by tidal stripping (Gnedin 1998) and ``starburst winds" as 
observed by Martin et al.\ (2002).  The metals are then assimilated 
into infalling clumps of IGM.  Details of this model remain unclear. 
Are the heavy elements expelled from galaxies in hot or warm gas?   How 
are the metals mixed with the IGM? What is the extent of their transport?  
Addressing these issues requires detecting and correlating many warm and hot 
IGM absorbers.

In this paper, we provide clear evidence for multi-phase IGM toward 
the BL Lac object PKS~2155-304.  This gas is associated 
with a small group of galaxies at $z \approx 0.0567$ and a cluster of 7 
strong Ly$\alpha$ absorbers (Shull et al.\ 1998) at velocities  
$cz = 16,283 - 17,570$ km~s$^{-1}$. This environment exhibits the diffuse 
warm phase (Ly$\alpha$, Ly$\beta$, Ly$\gamma$), the shocked (WHIM) phase 
(O~VI), and unconfirmed evidence (O~VIII) for much hotter, virialized gas 
residing at the bottom of the group's gravitational potential well.  

In \S~2, we describe our FUSE and HST observations of these absorbers in 
H~I and O~VI (and limits on C~IV, Si~III) 
and compare them to the claimed O~VIII absorption. 
In \S~3 we examine the absorber kinematics for the 
possibility of nearside/backside infall and ionization-state (O~VI/O~VIII) 
consistency.  We conclude in \S~4 with a discussion of an infall model 
that may be consistent with large-scale structure in this group.
In this picture, the H~I and O~VI absorbers arise from clumps of gas
falling at 200--400 km~s$^{-1}$ relative
to a much hotter ($10^{6.5-7.0}$~K) substrate that could be 
detectable in O~VII or O~VIII absorption. 
The metallicities may range from 0.02--0.1 solar.
All the phases predicted by simulations may be present in
this group, suggesting that clusters of strong Ly$\alpha$ and O~VI 
absorbers associated with groups of galaxies may be the 
``signposts" of shock-heated, metal-enriched baryons. 

\section{OBSERVATIONS}

In HST Cycle 9, we obtained new data from the STIS Echelle (E140M)  
with 10 orbits (PID 8125, 28.5 ksec) and compared the absorber velocities to 
those in a small group of galaxies containing four large H~I galaxies 
and one dwarf seen with the VLA (see Fig.\ 3 in Shull et al.\ 1998). 
The PKS~2155-304 sightline passes near the center of this group,
none of these galaxies lies closer than 400 kpc to the sightline, and no 
other dwarf galaxies were seen down to $m_B \approx 19$.  The galaxy $1D$ 
velocity dispersion is $\sigma_{\rm gal} = 325$ km~s$^{-1}$, and the Ly$\alpha$ 
velocity dispersion is $\sigma_{\rm Lya} = 740$ km~s$^{-1}$.  We estimate the
galaxy overdensity as $\delta \sim 100$, based on the group of five galaxies.  

We also observed PKS~2155-304 with FUSE (Moos et al.\ 2000)
in three visits on 23-24 October 1999 (P1080701, $t_{\rm exp} = 19.2$ ksec, 
and P1080705, $t_{\rm exp} = 38.6$ ksec) and 18 June 2001 (P1080703, 
$t_{\rm exp} = 65.4$ ksec). All data were calibrated with CALFUSE 2.2.1. 
We base most of our analysis on the second visit, which obtained data for 
all of the FUSE detectors, whereas the earlier visits occurred before the 
SiC and LiF channels (Sahnow et al.\ 2000) were aligned. For the LiF1a 
segment (1000--1084 \AA), we roughly tripled the amount of available data 
by co-adding night-only reductions of the three observations. This method 
recovered the redshifted Ly$\gamma$ absorption (1027.6 and 1028.0~\AA) from 
below O~I airglow emission (1025.76 and 1026.47 \AA). We applied small wavelength 
offsets ($\leq 0.05$ \AA) to the FUSE data to align narrow interstellar lines 
and place the FUSE and STIS data on a common velocity scale set by Galactic 
21-cm emission. 

The claimed X-ray detection of O~VIII absorption (Fang et al.\ 2002) 
with $4.5 \sigma$ significance was made with the {\it Chandra} Low-Energy
Transmission Grating Spectrometer. However, the redshifted O~VIII absorber
is not seen near 20~\AA\ in the spectrum taken by the {\it X-Ray Multi-Mirror
Observatory} (XMM) (see Fig. 1 and Table 1 in Rasmussen, Kahn, \& Paerels 2003).
Although the XMM velocity resolution is poorer than that of {\it Chandra}/LETG,
the XMM data have high S/N and detect oxygen absorption at $z = 0$ with
equivalent widths $16.3 \pm 3.3$ m\AA\ (O~VIII, 5.7 $\sigma$ significance) and
$9.0 \pm 2.7$ m\AA\ (O~VII, 8 $\sigma$ significance).  The non-confirmation of 
the $z = 0.0567$ O~VIII absorption remains a puzzle.
 
Figure 1 shows a velocity overlay of the Ly$\alpha$, Ly$\gamma$, O~VI, and
O~VIII absorbers between $cz = 15,000 - 20,000$ km~s$^{-1}$.
The detected lines and their properties are summarized in Table 1; 
the three strong Ly$\alpha$ absorbers are labeled A, B, and C for
clarity. We detect Ly$\beta$ at 16,243 km s$^{-1}$ (A), but Ly$\beta$ lines 
at 16,973 km~s$^{-1}$ (B) and 17,110 km s$^{-1}$ (C) are severely blended 
with the N~II $\lambda$1084 interstellar line. We detect Ly$\gamma$ in 
components B and C, but we derive a large error on their equivalent width
due to continuum placement uncertainty. We also detect O~VI in
components A and C, at 16,243 and 17,144 km~s$^{-1}$.

\section{INTERPRETATION OF THE SPECTRA}

With our new HST/STIS E140M data, we can derive more accurate column densities 
and metallicities for the strong Ly$\alpha$ absorbers.
In our previous study (Shull et al.\ 1998) at 20 km~s$^{-1}$ 
(HST/GHRS) resolution, we estimated N$_{\rm HI} = (3-10) \times 10^{14}$ cm$^{-2}$ 
for the blended B and C components at $z \approx 0.0567$.  There, we set a 
metallicity limit $Z < 0.003 Z_{\odot}$ from the absence of Si~III 
and C~IV absorption. At the higher S/N and better spectral resolution of the
STIS/E140M, together with FUSE measurements of Ly$\gamma$, the derived H~I columns 
decreased considerably. The integrated Ly$\alpha$ optical depths give 
N(H~I) = 1.0, 2.2, and $0.94 \times 10^{14}$ cm$^{-2}$, respectively,
for absorbers at 16,185, 16,974, and 17,147 km~s$^{-1}$, 
consistent with the observed Ly$\gamma$ strengths.

We still did not detect any Si~III $\lambda 1206$ or C~IV $\lambda 1548$ 
absorption at the Ly$\alpha$ velocities (Table 1). The 6--12~m\AA\ equivalent 
width limits correspond to N(C~IV) $< 2.9 \times 10^{12}$ cm$^{-2}$ and
N(Si~III) $< 2.7 \times 10^{11}$ cm$^{-2}$.  
As described in Shull et al. (1998), we assume that the absorption arises in a 
uniform slab of gas photoionized by the metagalactic radiation background,   
with spectral shape for a background dominated by AGN with intrinsic 
spectral index of $\alpha = 1.8$ modified by IGM absorption (Shull et al.\ 1999).  
The ionization correction (conversion from C~IV/H~I to [C/H]) depends only on 
the ionization parameter, $\log U \propto I_0 / n_H$, where $I_0$ is the 
specific intensity at 1 ryd and $n_H$ is the volume density of the gas.  
For $-2.0 < \log U < -0.5$, the metallicity upper limit from C~IV/H~I ranges 
between 0.025--0.10 of the solar abundance in the CLOUDY photoionization code, 
(C/H)$_{\odot} = 3.55 \times 10^{-4}$ (Grevesse \& Noels 1993). 
A full discussion of the parameter dependence of the metallicity 
estimate appears in a subsequent paper.

The two detected O~VI systems have column densities of N(O~VI) = $(2.8 \pm 0.5) 
\times 10^{13}$ cm$^{-2}$ (A) and $(3.3 \pm 0.5) \times 10^{13}$ cm$^{-2}$ (C)
assuming a linear curve of growth.  At the temperature (log $T_{\rm OVI} = 5.45$)
of maximum ionization abundance ($b_{\rm OVI} = 17$ km~s$^{-1}$), the stronger
\ion{O}{6} $\lambda 1032$ lines in both components would still be unsaturated.
We identify these two O~VI absorbers with 
Ly$\alpha$ systems at 16,185 (A) and 17,147 km~s$^{-1}$ (C).  No O~VI is seen 
in the 16,974 km~s$^{-1}$ (B) absorber.  The O~VI lines are the 
only metals yet detected in these absorbers, but their metallicity
is uncertain, owing to an indeterminate ionization state and thermal phase 
(see Savage et al.\ 2002).  If the O~VI arises in photoionized gas, its column 
density and limits on C~IV imply $\log U > -0.8$, for C/O in a solar abundance 
ratio of $0.50 \pm 0.07$ (Allende Prieto, Lambert, \& Asplund 2002).

If the ionizing background has $I_0 \approx 10^{-23}$ 
ergs~cm$^{-2}$ s$^{-1}$ Hz$^{-1}$ sr$^{-1}$ (Shull et al.\ 1999), the 
density constraint is $n_H < 10^{-5.6}$ cm$^{-3}$, and the corresponding 
line-of-sight extent of the gas is $D > 1.1$~Mpc, comparable to the
extent of the galaxy group.  We believe that this low-density, large-absorber
model is highly unlikely.  Multiple absorbers of this extent could not fit 
within the 1 Mpc volume of the group, without undergoing collisions and
shredding.  However, if the O~VI arises in collisionally ionized gas, the 
requirement that the gas be so rarefied no longer holds.  
In addition, the fraction of O~VI in collisional 
ionization equilibrium peaks at $T \sim 10^{5.5}$ K, and it is tempting to 
associate the O~VI with the WHIM.  The tentative detection of O~VIII absorption 
makes collisionally ionized O~VI still more compelling.
The claimed strength of the X-ray feature (Fang et al.\ 2002) requires 
log N(O~VIII) $\approx 16.0 \pm 0.2$, which could represent hot gas at the 
group barycenter. However, redshifted O~VIII absorption was not confirmed by 
XMM, nor has O~VII absorption at $z = 0.0567$ been detected 
(Nicastro et al.\ 2002).

\section{DISCUSSION}

In order to understand the cluster of PKS~2155-304 absorbers at 
$z \approx 0.0567$, one must resort to a multiphase model. The H~I, O~VI, 
and O~VIII systems are unlikely to co-exist, owing 
to kinematic offsets and physical considerations.  For example, the
claimed O~VIII absorption centroid ($16,624 \pm 237$ km~s$^{-1}$) 
appears to differ from the observed H~I and O~VI absorbers (A, B, C 
components).  Also, the observed Ly$\gamma$ line widths 
(FWHM $\leq 30$ km~s$^{-1}$) rule out H~I temperatures greater than 
20,000~K.   The measured O~VI line widths ($40 \pm 10$ km~s$^{-1}$ FWHM) 
make it unlikely, although marginally possible within errors, for O~VI 
and O~VIII to exist at the same temperature, $\log T \approx 6.25 \pm 0.1$,
where (O~VI/O~VIII) $\approx 6 \times 10^{-3}$ in collisional
ionization equilibrium (Sutherland \& Dopita 1993).  However, we 
discount this scenario, because of the observed velocity offsets
($\sim 400$ km~s$^{-1}$).  For similar reasons, the strong H~I absorbers 
cannot co-exist with the hot O~VIII.  At the high temperatures 
needed to ionize O~VIII collisionally, the H~I ionization fraction
would be too small to be detected.

Therefore, we favor a more complex scenario in which the H~I and O~VI
absorbers arise in clumps of gas falling into a small-group potential. 
Any hot gas, visible in O~VII or O~VIII, would then exist at the barycenter
of the group (16,600 km~s$^{-1}$) which would be consistent
with the kinematic offsets.  If this model is correct, the two O~VI
absorbers (A and C) would arise from clumps
undergoing ``backside" and ``nearside" infall, respectively.  Because
production of collisionally ionized O~VI requires shock velocities 
$V_s \geq 130$ km~s$^{-1}$ (Shull \& McKee 1979; Dopita \& Sutherland 1996),
the infall model has consequences for which Ly$\alpha$ absorbers
contain associated O~VI.  In particular, the absence of detectable
O~VI in the strongest Ly$\alpha$ system (B) may be
a result of a low relative velocity between the infalling H~I clump
and the substrate.  From Figure 1, one sees that O~VI is
present in the H~I absorber at 17,147 km~s$^{-1}$, located 
approximately 170 km~s$^{-1}$ redward.  Thus, one might speculate
that these two blended Ly$\alpha$ absorbers, one with and one without 
O~VI, represent shocked and unshocked gas.

In the clump-infall model, the relative velocity between the two O~VI
components and the group barycenter is $\sim 400$ km~s$^{-1}$, which
would produce a post-shock temperature of $\sim 2 \times 10^6$~K,   
similar to that ($2.5 \times 10^6$~K)
at which the observed O~VI and claimed O~VIII would co-exist in
collisional ionization equilibrium.  The cooling time of such gas is
\begin{equation}
   t_{\rm cool} = \frac {(20~{\rm Gyr}) T_{6.5}} {n_{-4} \Lambda_{-23}} \; , 
\end{equation}
where $n_{-4} = n_H/(10^{-4}~{\rm cm}^{-3})$ and $T_{6.5} = (T / 10^{6.5}$~K). 
Here, $\Lambda_{-23}$ is the radiative cooling rate coefficient in units of 
$10^{-23}$ erg~cm$^3$~s$^{-1}$, typical of gas at $10^{6.5}$~K 
with 0.1 solar metallicity 
(Sutherland \& Dopita 1993). In the range $5.0 < \log T < 7.0$, 
$\Lambda(T) \propto T^{-1}$, so $t_{\rm cool}$ is longer 
than the age of the universe, unless the gas temperature can
be decreased ($T_{6.5} \ll 1$) or the density increased ($n_{-4} \gg 1$).   
Through shock-compression and cooling, the cooling will 
accelerate as $\Lambda(T)$ moves toward its $10^{5.3}$~K peak. 

We can relate the hydrogen density to the cosmological overdensity 
$\delta$ by: 
\begin{equation}
    n_H(z) = (1.90 \times 10^{-7}~{\rm cm}^{-3})(1+z)^3
       \left[ \frac {\Omega_b h_{70}^2}{0.046} \right] (1 + \delta)  \; . 
\end{equation}
At the redshift ($z_a = 0.0567$) of the strong absorbers, we have 
$n_{-4} \approx (\delta / 500)$.  If the infalling gas is shocked 
to $T_s = (1.24 \times 10^6~{\rm K})(V_s/300~{\rm km~s}^{-1})^2$
to form the WHIM phase, it must acquire a fairly high density
($\delta > 500$ at $z \approx 0.1$) in order to cool, recombine, and form 
the observed O~VI. Adiabatic shocks can provide a compressive trigger, but
further radiative cooling is required, since adiabatic compression
lengthens the cooling time ($t_{\rm cool} \propto n^{1/3}$ for
$T \propto n^{2/3}$ and $\Lambda(T) \propto T^{-1}$).   Lower-velocity shocks, 
$V_s \approx 200$ km~s$^{-1}$, would cool more rapidly, with temperatures 
near the peak of the cooling curve.  

Two additional issues concern the ratios of C~IV, O~VI, and H~I.  
In the two detected O~VI absorbers, (C~IV/O~VI) $< 10^{-1.1}$, 
whereas models of radiatively cooling gas (Dopita \& Sutherland 
1996; Indebetouw \& Shull 2003) find (C~IV/O~VI) $\approx 0.1-0.2$.  
{\it Where is the C~IV?}  Is the cooled, recombined layer truncated in the 
hot-gas environment?  Similarly, we do not understand the wide variation in 
[N(H~I)/N(O~VI)] $\approx$ 0.1--10 among other IGM absorbers (Shull 2003). 
The production of shocked O~VI requires 
{\it relative} velocities above $\sim 150$ km~s$^{-1}$ and O/H abundances
above a few percent solar. However, the production mechanism of O~VI remains 
unclear (Sembach et al. 2003). While shock-heating is the preferred mechanism, 
other processes have been proposed involving interfaces between hot substrates 
and moving clouds -- thermal conduction, shear instabilities, and turbulent
mixing (see Indebetouw \& Shull 2003 for a review).  These
processes all have problems explaining the observed components,
N(O~VI) $\approx 3 \times 10^{13}$ cm$^{-2}$.   Each conductive 
interface typically produces N(O~VI) $\sim 10^{13}$ cm$^{-2}$, while
shock heating requires supersonic flow and depends on the Mach number cubed.  
If $T_{\rm hot} \approx 10^{6.4}$~K (adiabatic sound speed 
$c_s \approx 240$ km/s), infalling clumps at 300-400 km~s$^{-1}$ have Mach 
numbers of just 1.25 -- 1.67.  

Resolving the O~VI heating paradox may require understanding the cloud-substrate 
interfaces at a deeper level.  Alternatively, if the O~VIII absorption
turns out to be non-existent, a virialized hot substrate may not have formed
yet. Intergroup gas at $T < 10^6$~K would produce sufficient shock-heating,
and the observed O~VI absorption may arise from the interactions of infalling
clumps with metals in a intergroup medium injected from the
galaxies.  With combined UV and X-ray spectra, one may be able to distinguish among  
these processes through the amount of shock heating and the relative abundances 
of O~VI, O~VII, and O~VIII.

\acknowledgements

This work is based on data obtained for the Guaranteed Time Team by the 
NASA-CNES-CSA FUSE mission operated by the Johns Hopkins University. 
Financial support to U.S. participants has been provided by
NASA contract NAS5-32985. We were also supported by
grants GO-08571.01-A from the Space Telescope Science Institute and
NAG5-7262 from NASA/LTSA.  We thank Taotao Fang for providing the 
{\it Chandra} spectrum that appears in Figure 1 and John Stocke 
and Bill Blair for helpful discussions.

\newpage

\begin{deluxetable}{lccc}
\tablecolumns{4}
\tablewidth{0pc}
\tablecaption{PKS~2155-304 IGM Absorption Lines}
\tablehead{
\colhead{Line} & \colhead{Velocity}     & \colhead{$W_{\lambda}$}
              &  \colhead{$\Delta V$\tablenotemark{a} }  \\
\colhead{}    & \colhead{(km~s$^{-1}$)} & \colhead{(m\AA)} &
                \colhead{(km~s$^{-1}$)} }
\startdata
  H I Ly$\alpha$ (A)   & $16185\pm 5$    & $331 \pm 12$      & $75 \pm 8$  \\
  H I Ly$\alpha$ (B)   & $16974\pm 5$    & $495 \pm 10$      & $76 \pm 5$  \\
  H I Ly$\alpha$ (C)   & $17147\pm 5$    & $353 \pm 8$       & $110 \pm 10$  \\
  H I Ly$\beta$  (A)   & $16243\pm 10$   & 85 $\pm$ 10       & $92 \pm 17$  \\
  H I Ly$\gamma$ (B)   & $16936\pm 10$   & 16 $\pm$ 4        & $<30$        \\
  H I Ly$\gamma$ (C)   & $17109\pm 10$   & 13 $\pm$ 3        & $<30$       \\
  H I Ly$\delta$\tablenotemark{b} & \nodata  & $<$12         & $<30$       \\
  O VI 1032 (A)        & $16243\pm 10$   & 37 $\pm$ 7        & $35 \pm 10$  \\
  O VI 1038 (A)        & $16252\pm 10$   & 25 $\pm$ 5        & $30 \pm 10$   \\
  O VI 1032 (C)        & $17144\pm 10$   & 44 $\pm$ 6        & $40 \pm 10$   \\
  O VI 1038 (C)        & $17116\pm 10$   & Fe II blend       &  ...          \\
  O~VIII 18.969        & $16624\pm 237$  & $14.0^{+7.3}_{-5.6}$  &  $<1380$   \\
  C III  977\tablenotemark{b}  & \nodata &  $<10$    &  \\
  C IV 1548\tablenotemark{b}   & \nodata &  $<12$    &  \\
  Si III 1206\tablenotemark{b} & \nodata &  $<6$     &  \\
\enddata
\tablenotetext{a}{Full width at half maximum (FWHM) from STIS/FUSE 
   profile fitting, or by converting {\it Chandra} line width of
   $\sigma({\rm O~VIII)} < 0.039$~\AA\ (Fang et al.\ 2002).  The doppler 
   parameter, $b_{\rm dopp} = {\rm FWHM}/[2 (\ln 2)^{1/2} ]$. }

\tablenotetext{b}{Upper limit at 4$\sigma$ significance.}
\end{deluxetable}

\newpage

\begin{figure*}
 \centerline{\epsfxsize=0.55\hsize{\epsfbox{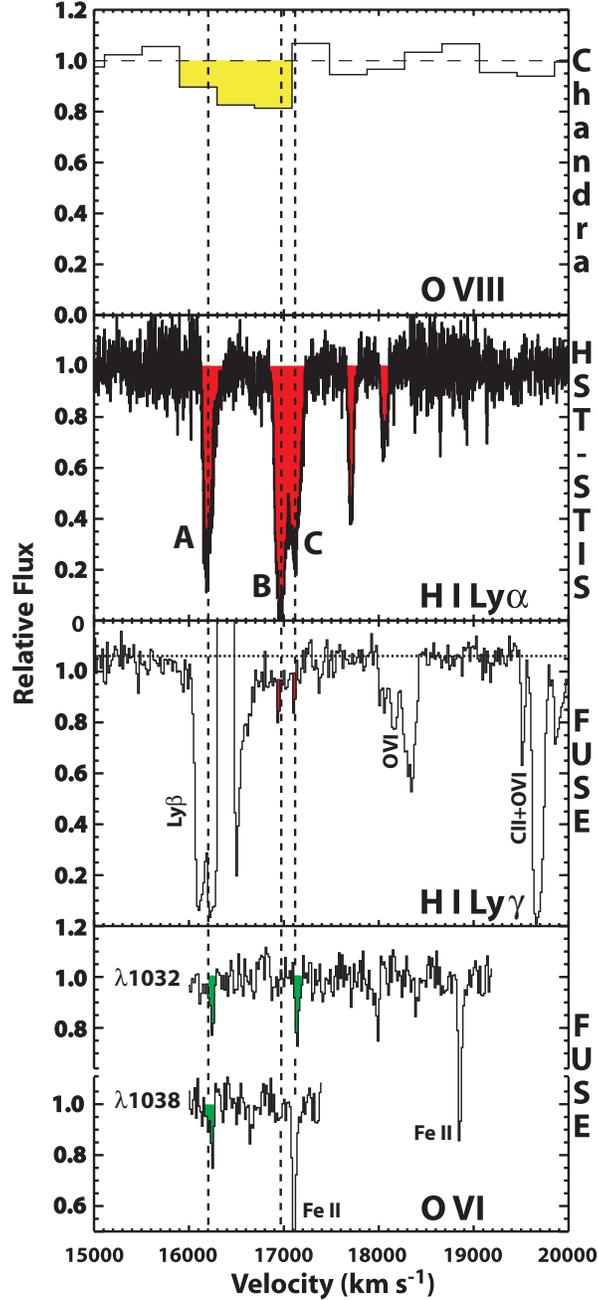}}}
 \figcaption{Absorption lines of H~I (Ly$\alpha$ from HST/STIS, Ly$\gamma$
  from FUSE), O~VI 1032/1038 (FUSE), and claimed detection of O~VIII
  with {\it Chandra} (Fang et al.~2002) toward PKS~2155-304.  The three
  strongest Ly$\alpha$ absorbers are labeled (A,B,C -- see Table 1). We
  detect O~VI in the 16,243 km~s$^{-1}$ (A) and 17,144 km~s$^{-1}$ (C)
  absorbers, and Ly$\gamma$ in B and C absorbers; claimed O~VIII absorption
  is offset by $\sim400$ km~s$^{-1}$ from Ly$\alpha$ and O~VI
  (vertical dotted lines). Interstellar lines (Ly$\beta$, C~II, O~VI)
  are labeled in panel 3.  Lower curve in panel 4 shows O~VI~1038
  offset from O~VI 1032.  The Fe~II line shows an asymmetric profile
  that suggests presence of O~VI~1038 at 17,140 km~s$^{-1}$, but blending
  precludes an accurate measurement.}
\label{stackfig}
\end{figure*}
\vspace{0.1in}

\end{document}